\documentclass[final,3p,times,twocolumn]{elsarticle}

\usepackage{amssymb}
\usepackage{graphicx}

\journal{Journal of Crystal Growth}

\begin{document}

\begin{frontmatter}



\title {Thermodynamic modeling of the LiF$-$YF$_3$ phase diagram}


\author[IPEN]{I. A. dos Santos\corref{cor1}}
\cortext[cor1]{corresponding author}
\ead{iasantosif@yahoo.com.br}

\author[IKZ]{D. Klimm}
\author[IPEN]{S. L. Baldochi}
\author[IPEN]{I. M. Ranieri}

\address[IPEN]{Instituto de Pesquisas Energ\'eticas e Nucleares, CP 11049, Butant\~a 05422-970, S\~ao Paulo, SP, Brazil}
\address[IKZ]{ Leibniz Institute for Crystal Growth, Max-Born-Stra\ss e 2, 12489 Berlin, Germany}

\begin{abstract}
A thermodynamic optimization of the LiF$-$YF$_3$ binary phase diagram was performed by fitting the Gibbs energy functions to experimental data that were taken from the literature, as well as from own thermoanalytic measurements (DTA and DSC) on HF-treated samples. The Gibbs energy functions for the end member compounds were taken from the literature. Excess energy terms, which describe the effect of interaction between the two fluoride compounds in the liquid phase, were expressed by the Redlich-Kister polynomial function. The calculated phase diagram and thermodynamic properties for the unique formed compound, LiYF$_4$, are in reasonable agreement with the experimental data. 
\end{abstract}

\begin{keyword}
 A1. Phase diagrams \sep A1. Computer simulation \sep A1. Characterization \sep B1. Rare earth compounds


\end{keyword}

\end{frontmatter}


\section{Introduction}
\label{sec:intro}

The main interest of the system LiF$-$YF$_3$ is due to the technological applications of LiYF$_4$ crystals (YLF) as active laser medium \cite{Ranieri04,Wetter00}. Thoma \textit{et al.} determined the phase diagram of this system using differential thermal analysis (DTA) data. A eutectic reaction was observed for the composition of 80\,mol\% LiF$-$20\,mol\% YF$_3$ at 963\,K and a peritectic reaction for the composition of 49\,mol\% LiF$-$51\,mol\% YF$_3$ at 1098\,K \cite{thoma02}.

Harris \textit{et al.} \cite{Harris83} performed an extensive study of the melting behavior of YLF by DTA and microstructural studies, it was assumed that YLF has a congruent melting behavior. Nevertheless in the growth of crystals by the Czochralski method, YLF was easily seeded only with the composition of 50.7\,mol\%LiF$-$49.3\,mol\%YF$_3$. They concluded that the degree of congruence of this compound is closely related to contamination with oxygen compounds, mainly with moisture \cite{Safi81,Abell76}. Then it was proposed a phase diagram of the system LiF$-$YF$_3$ composed by two eutectics, one with composition of 80\,mol\% LiF$-$20\,mol\% YF$_3$ at 979(2)\,K, and the other to the composition 49\,mol\% LiF$-$51\,mol\% YF$_3$ at temperature 1103(2)\,K, the same temperature was attributed to the melting of YLF.

Although the experimental phase diagram of this system is well known, there is scarce data concerning thermodynamic properties of fluorides. Regarding the rare earth trifluorides, including YF$_3$, Barin \cite{Barin93} collected the commonly used data. Lyapunov \textit{et al.} obtained YLF enthalpies and heat capacities in the solid and liquid state by the mixing method using a massive calorimeter \cite{Lyapunov00}.

In this work, the LiF$-$YF$_3$ phase diagram was optimized by the fitting of the Gibbs energy function, taking into account both the experimental data from the literature \cite{Barin93} and the data obtained in this work using the differential thermal analysis (DTA). Gibbs excess energy terms in the liquid phase, which describe the effects of interaction between the two fluorides, were expressed by the Redlich-Kister polynomial function \cite{Redlich00}. Heat capacity ($C_p$), enthalpy of formation and entropy at 298.15\,K were assessed for the YLF compound and compared with those reported in the literature.

\section{Experimental}
\label{sec:exp}

Samples were prepared using commercial LiF (Ald\-rich, 99.9\%) purified by the zone melting method, and YF$_3$ synthesized from the oxide (Y$_2$O$_3$, Alfa Aesar, 99.99\%) by the hydrofluorination method \cite{Guggenheim63}, under reactive atmosphere of HF and Ar in both cases. DTA curves were obtained using a TGA-DTA equipment from TA Instruments, model 2960. The experiments were performed under Ar flow, using Pt/Au crucibles; heating rate of 10\,K/min and samples masses around 50\,mg. The melting point of the pure substances, the temperature of the solid phase transformation and of the invariant reactions were considered as the extrapolated onset of the thermal event. The liquidus temperatures of the intermediary compositions were evaluated from the extrapolated offset temperatures.

The heat capacity of the YLF was evaluated using a Netzsch STA 409 PC Luxx heat-flux differential scanning calorimeter (DSC), where a proper $C_P$ carrier was installed. DSC setup was temperature calibrated measuring the melting points of Zn, Au, In, Ni, and the phase transformation of BaCO$_3$. All experiments were carried out under Ar flow of 50\,cm$^3$/min and the samples were place in Pt/Au crucibles with lid. YLF heat capacity was determined according to ASTM-E-1269 method. In this method, the sample DSC heat flow signal is compared to the DSC signal of a calibration standard of known specific heat (sapphire in this case). Both curves are corrected by a baseline correction experiment where empty reference and sample crucibles are placed in the DSC furnace and the system signal drift is measured under identical experimental conditions. The experiments consisted of three steps, an isothermal segment in 40$^{\,\circ}$C for 20\,min, a dynamic heating segment with a 10\,K/min heating rate and a final isothermal segment at maximum temperature for 5\,min.

\section{Thermodynamic method}
\label{Thermo}

To describe a $T - X$ binary phase diagram it is necessary to define the Gibbs energy functions for all compounds in the system and the Gibbs functions of mixing, if solution phases are present. Usually these functions are unknown for most of the solutions, thus a thermodynamic assessment is required in order to determine the excess energy of the mixing. 
The Gibbs equation for the compounds is defined as function of enthalpy and entropy at the reference temperature state (298.15\,K) and can be obtained from the heat capacity function ($C_P(T)$) as follows:

\begin{equation}
G(T)=H(T)-S(T)T
\label{eq1} 
\end{equation}

\begin{equation}
H(T)=H^{0}_{298.15\,\mathrm{K}}+\int_{298.15\,\mathrm{K}}^{T}C_P \,\mathrm{d}T
\label{eq2} 
\end{equation}

\begin{equation}
S(T)=S^{0}_{298.15\,\mathrm{K}}+\int_{298.15\,\mathrm{K}}^{T}\frac{C_P}{T} \,\mathrm{d}T
\label{eq3} 
\end{equation}

The $C_P(T)$ functions can be obtained by fitting a set of experimental data at a suitable polynomial function expressed by:

\begin{equation}
C_P(T)=a+bT+cT^{-2}
\label{eq4} 
\end{equation}

Depending on the compound, more terms in this $C_P(T)$ expression are added or disregarded in order to obtain the best fitting.
The thermodynamic data for the end members LiF and YF$_3$ were taken from the compilation by Barin \cite{Barin93}. According to equation (\ref{eq1}) and considering the equations (\ref{eq2}) and (\ref{eq3}), the enthalpy of formation, the absolute entropy at the reference temperature and the heat capacity are required to determine the minimum of $G$ and thus thermodynamic equilibrium. For the LiYF$_4$ intermediate compound these data are not available, therefore the Neumann-Kopp rule was assumed to set the initial $C_P$ equation and calorimetric data, afterwards these latter parameters were properly assessed by optimization.

\begin{table*}[ht]
\renewcommand\footnoterule{}
\caption{$\Delta H${(298.15\,K)} (kJ\,mol$^{-1}$), $S${(298.15\,K)} (J\,K$^{-1}$\,mol$^{-1}$), $\Delta H_f$ (kJ\,mol$^{-1}$) and $C_P$ data for LiF, YF$_3$ and the intermediate compound LiYF$_4$.}
	  \begin{minipage}{14cm}
	  \renewcommand{\arraystretch}{1.1}
	  \centering
	  \begin{tabular}{crrrrrrrr}
		\hline
Compound   &  $\Delta H${(298.15\,$K$)}   &   $S${(298.15\,$K$)}    &  $\Delta H_{f(1)}$ \footnote{Data taken from Barin \cite{Barin93} and from Lyapunov \textit{et al.} \cite{Lyapunov00}.} & $\Delta H_{f(2)}$ \footnote{DSC data measured in this work and from \cite{Klimm08b}.}&    $a$      &      $b$     & $c$ \\
\hline
LiF$(S)$   &  $-616.931$   &    35.660   &   27.09 \cite{Barin93}   &   27.68  &  42.689   &    $1.742\times10^{-2}$ & $-5.301\times10^{5}$\\
LiF$(l)$   &  $-594.581$   &    42.997   &     --                   &     --    &  64.183   &      --        &      --     \\
YF$_3(S1)$ &  $-1718.368$  &    109.960  &     --        &     --    &  99.411   & $7.427\times10^{-3}$ & $-5.690\times10^{5}$\\
YF$_3(S2)$ &  $-2161.269$  &   -960.528\footnote{Extrapolated value for the high temperature phase.}  &   27.97 \cite{Barin93}   &   29.79 \cite{Klimm08b}  & $-319.448$  &    $2.128\times10^{-1}$    &     $2.818\times10^8$\\         
LiYF$_4$\footnote{$\Delta H${(298.15\,$K$)} and $S${(298.15\,$K$)} were assessed in this work and $C_P$ function was estimated by the Neumann-Kopp rule.}   & $-2355.780$  &   138.325  &   67.65 \cite{Lyapunov00}   &   63.538     &   142.101   &   $2.484\times10^{-2}$  &  $-1.099\times10^6$  \\
\hline
		\end{tabular}
		\end{minipage}
   	\label{table01}
	  \end{table*}

For the liquid solution phase, function $G(T)$ is expressed as the sum of the Gibbs energy weighed contribution of the pure compounds ($G_0$), the contribution of an ideal mixture ($G_{ID}$) and finally a term related to the non-ideal interaction, defined as the excess energy ($G_{ex}$). The sub-regular solution model of Redlich-Kister was adopted to describe the excess energy of the liquid phase in this system \cite{Redlich00}. For binary systems, the polynomial model is expressed by:
\begin{equation}
G_{ex}=x_Ax_B\sum_{j=0}^{N}L_j\,(x_A-x_B)^j
\label{eq5} 
\end{equation}
where $x_A$ and $x_B$ are the molar fractions of components $A$ and $B$ , respectively. $L_j$ terms represent the interaction coefficients between the basis compounds and they are given as a linear function of temperature. The optimization was performed using the OptiSage module in the FactSage 6.2 software \cite{FactSage6_2}, which uses the Bayesian Algorithm \cite{Pelikan00}. This algorithm is based on a probability model to obtain the fit between the theoretical Gibbs energy functions and the experimental data. 

\section{Results and discussion}
\label{sec:results}

\subsection{Experimental results}

\begin{figure}[htb]
\centering
\includegraphics[width=0.44\textwidth]{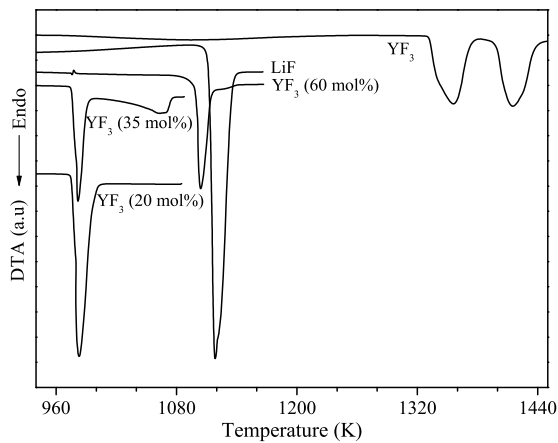}
\caption{DTA curves for several LiF--YF$_3$ compositions.}
\label{fig1}
\end{figure}

DTA curves for pure LiF, for three mixed LiF--YF$_3$ samples with different composition, and for pure YF$_3$ are shown in Figure~\ref{fig1}. The curves for LiF and LiF+20\,mol\%\,YF$_3$ exhibit only one endothermic peak with onsets near 1115 or 975\,K, respectively. These peaks are due to LiF melting ($\approx1115$\,K) and the LiF/LiYF$_4$ eutectic ($\approx975$\,K, the eutectic point is close to this composition, \emph{cf.} Figure~\ref{fig3}). DTA curves for 35 and 60\,mol\%\,YF$_3$ show two endothermic events, the first peak characterizes an invariant reaction (\textit{e.g.} eutectic or peritectic reaction) and the second broader peak marks the end of melting of the primary phase (YLF and YF$_3$ respectively) at the liquidus. In the YF$_3$ curve the solid state transformation (1338\,K) and fusion (1403\,K) can be recognized very clearly. These two thermal events show very similar peaks in area, therefore similar enthalpy is required in these transformations, as it was observed previously \cite{Klimm08b}.

\begin{figure}[htb]
\centering
\includegraphics[width=0.45\textwidth]{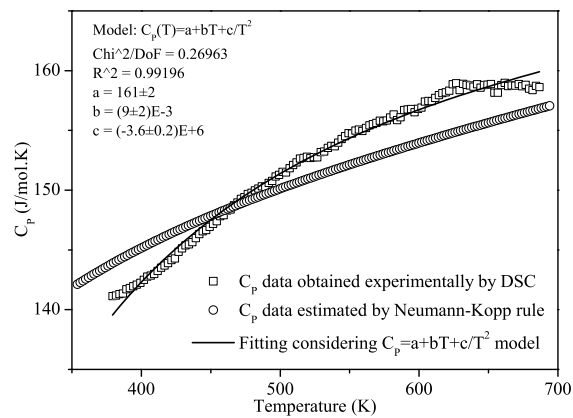}
\caption{$C_P$ data measured experimentally through DSC technique and $C_P$ estimated by Neumann-Kopp rule.}
\label{fig2}
\end{figure}

The heat of fusion $\Delta H_{f}$ for LiF, YF$_3$ and LiYF$_4$ was calculated from the melting peak area of these compounds. Reasonable agreement with the literature was found in the $\Delta H_{f}$ (Table~\ref{table01}). Nevertheless  the Neumann-Kopp rule had been considered to set YLF $C_P(T)$ function used to perform the assessment, $C_P$ data for this intermediate compound were also measured by DSC. Figure~\ref{fig2} compares DSC measured data and those estimated by the Neumann-Kopp rule. One can see that for the considered temperature range, there are no strong deviations (not bigger than 3\%) between the experimental and estimated $C_P$ data. Moreover, taking into account the polynomial fitting of the DSC experimental data using the function (\ref{eq4}), the $a$ and $b$ parameters obtained are very similar to those achieved by Lyapunov \textit{et al.} \cite{Lyapunov00}, and the difference is less than 3\% for the $c$ parameter.

\subsection{Thermodynamic assessment}

\begin{figure}[htb]
\centering
\includegraphics[width=0.47\textwidth]{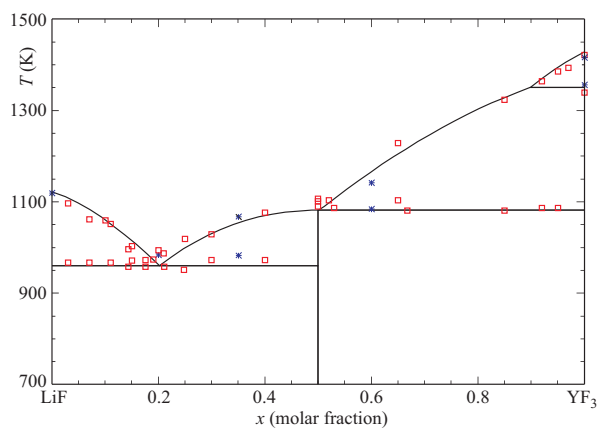}
\caption{Theoretical phase diagram of the system LiF$-$YF$_3$ calculated assuming the experimental data from the literature (squares) and those measured in this work (stars).}
\label{fig3}
\end{figure}

A proper optimization of binary systems using a polynomial model to represent the Gibbs energy for the solution phases is strongly dependent on quantity and accurate of experimental data available for the system studied. Therefore, the experimental phase diagram of the system LiF$-$YF$_3$ was constructed taking into account the experimental data evaluated by Thoma \textit{et al.} \cite{thoma02} (squares), and the DTA data obtained in this work (stars) (Figure \ref{fig3}). It can be seen that our experimental points are in agreement with those collected from literature. 

The LiF$-$YF$_3$ phase diagram has been optimized according to the Redlich-Kister polynomial model using the Bayesian Optimization Algorithm of FactSage \cite{FactSage6_2}. The data available in the literature for this system are data of solidus and liquidus lines (experimental $T - X$ phase diagram) and $C_P$ and calorimetric properties for the end members (LiF and YF$_3$). No enthalpies of mixing or activity data were considered for the LiF$-$YF$_3$ liquid phase. Consequently many degrees of freedom are present in the assessment, making the setting of the appropriately solution features not straightforward. It was observed by other authors working with different LiF--$Ln$F$_3$ systems ($Ln$ = La--Sm; these systems do not contain a Li$Ln$F$_4$ phase), that the optimized Redlich-Kister coefficients for one and the same system can be different \cite{Meer00}. 

Figure~\ref{fig3} shows the LiF$-$YF$_3$ optimized phase diagram that was obtained in this work, together with the $T - X$ experimental points for comparison. It may be noticed that the assessed values of the eutectic reaction temperature and the melting temperature of LiYF$_4$ are in good agreement with experimental data. The assessed enthalpy and entropy at 298.15\,K temperature are listed in table~\ref{table01}. The excess parameters assessed, given by $L_0$ and $L_1$ in the Redlich-Kister model (equation~\ref{eq5}) were $L_0=-15883.80-24.18T$ and $L_1=42271.89-23.26T$.

\section{Conclusions}

Thermodynamic assessment has been performed on the LiF$-$YF$_3$ binary system. The excess Gibbs parameters for the liquid phase could be properly optimized using a Redlich-Kister polynomial model. The resulting theoretical phase diagram shows satisfactory agreement compared to the experimental phase diagram. $\Delta H${(298.15\,K)} and $S${(298.15\,K) have been assessed for LiYF$_4$, and a re-evaluation for $\Delta H_f$ and $C_P$ data for this intermediate compound was performed. It should be noted that the liquidus from the YF$_3$ rich side meets the LiYF$_4$ liquidus at a molar fraction of 50\%\ YF$_3$. Results presented in this paper are contributions to the more complete thermodynamic description of LiF$-$YF$_3$ system and may be useful for the further thermodynamic assessment, in particular for ternary and multi component systems based on this binary phase diagram.

\section*{Acknowledgments}

The authors acknowledge financial support from CNPq (477595/2008-1; 290111/2010-2) and DAAD-CAPES (po-50752632).











\end{document}